\documentclass{vgtc} 

\newif\ifnotes
\notesfalse 

\newif\iftrack
\trackfalse 

\onlineid{1004}

\vgtccategory{Research}

\vgtcpapertype{evaluation}

\usepackage{microtype}                 
\PassOptionsToPackage{warn}{textcomp}  
\usepackage{textcomp}                  
\usepackage{mathptmx}                  
\usepackage{times}                     
\usepackage{ccicons}
\usepackage{booktabs}                  
\usepackage{float}
\usepackage{dblfloatfix} 
\usepackage{lettrine}
\usepackage{csquotes}

\usepackage[normalem]{ulem}

\newcommand{\app}{Tableau for visionOS}

\title{Evaluating an Immersive Analytics Application at an \\Enterprise Business Intelligence Customer Conference}

\author{
    Matthew Brehmer$^\dagger$\thanks{e-mail: mbrehmer@uwaterloo.ca}\\ %
     \scriptsize University of Waterloo, Waterloo ON Canada %
\and Ginger Gloystein\\ %
     \scriptsize Tableau, Seattle WA USA %
\and Bailiang Zhou\\ %
     \scriptsize Tableau, Seattle WA USA %
\and Abby Gray\thanks{work performed at Tableau}\\ %
     \scriptsize Cedar AI, Seattle WA USA %
\and Sruthi Pillai\\ %
     \scriptsize Tableau, Seattle WA USA %
\and Ben Medina$^\dagger$\\ %
     \scriptsize Adobe, Seattle WA USA %
\and Vidya Setlur\thanks{e-mail: vsetlur@tableau.com}\\ %
    \scriptsize Tableau, Palo Alto CA USA}

\abstract{%
    We reflect on an evaluation of an immersive analytics application (Tableau for visionOS) conducted at a large enterprise business intelligence (BI) conference. Conducting a study in such a context offered an opportunistic setting to gather diverse feedback. However, this setting also highlighted the challenge of evaluating usability while also assessing potential utility, as feedback straddled between the novelty of the experience and the practicality of the application in participants' analytical workflows. This formative evaluation with 22 participants allowed us to gather insights with respect to the usability of Tableau for visionOS, along with broader perspectives on the potential for head-mounted displays (HMDs) to promote new ways to engage with BI data. Our experience suggests a need for new evaluation considerations that integrate qualitative and quantitative measures and account for unique interaction patterns with 3D representations and interfaces accessible via an HMD. Overall, we contribute an enterprise perspective on evaluation methodologies for immersive analytics.
}


\graphicspath{{figs/}{figures/}{pictures/}{images/}{./}} 

\usepackage{tabu}                      
\usepackage{booktabs}                  
\usepackage{lipsum}                    
\usepackage{mwe}                       

\usepackage{hyperref}
\usepackage[capitalise]{cleveref}

\usepackage{mathptmx}                  

\begin{document}



\maketitle

\section{Introduction}

The emergence of head-mounted displays (HMDs) has transformed the landscape of interactive experiences. Devices such as Apple's Vision Pro~\cite{visionpro}, HTC's Vive~\cite{vive}, Meta's Quest~\cite{metaquest}, and Microsoft's HoloLens~\cite{hololens} have moved beyond niche gaming and entertainment platforms to become promising platforms for knowledge work, as demonstrated by the growing interest in \textit{immersive analytics}~\cite{saffo2023unraveling}. 
They have the potential to manipulate virtual data objects in novel ways, thereby resulting in a deeper level of immersion~\cite{ens2021grand}. 

The potential of immersive analytics elicits two major evaluation objectives. 
First, we must better understand how to design applications that are both \textit{engaging} and \textit{useful}, particularly in the domain of enterprise BI, where until recently there have been few commercial application offerings for HMDs and a general unfamiliarity with these devices.
Second, we need to identify best practices for maintaining information clarity and usability given the dynamic conditions typical of immersive environments. 

The primary contribution of our work is a reflection on a recent evaluation of an HMD-based immersive analytics experience for the BI domain.
At a large enterprise business intelligence conference, we conducted an opportunistic formative evaluation of \app{}, a video-see-through (VST) augmented reality (AR) BI application built for the Apple Vision Pro. 
Based on observations and conversations with 22 participants, we arrived at insights relating to the novelty, usability, and potential utility of our application. 
While participants found the interaction paradigms in our immersive analytics experience to be novel, there was a notable dichotomy in how expert BI practitioners interacted with \app{} relative to novices; the former adapted quickly to an immersive analytics workflow, while novices struggled and required additional guidance. 

In reflecting on our observations, we identified a tension between usability and the elicitation of use cases for immersive BI tasks, highlighting the challenges inherent in technology-driven innovation and understanding its ecological validity.  
This tension is particularly pronounced where novel interaction modalities such as gaze and gesture-based controls could be both exciting and perplexing for users. 
Our application introduced interfaces that often do not align with users' existing mental models or analytics workflows, suggesting a possible learning curve that can impact application learnability and usability. 
We further acknowledge a tension between participants' excitement with respect to a novel technology medium and their ability to articulate or envision the applicability of this technology in a real BI context, particularly given the difference between their work environment and our study setting. 
We distill our reflection into a set of recommendations for confronting and mitigating these tensions in future evaluations of immersive analytics applications and experiences that go beyond considerations of usability and performance on well-defined tasks.

\section{Related Work}

Evaluations that consider both usability and utility are crucial for understanding the full potential of immersive analytics. 
Friedl-Knirsch et al.'s recent systematic literature review of user evaluation in immersive analytics summarizes the methodologies used in prior work and identifies underrepresented study designs~\cite{Friedl-Knirsch2024}. 
Their survey considers a wide range of evaluation goals, from prototype evaluation to foundational research, and highlights the need for a shared evaluation framework to further research in immersive analytics. 
This dual focus on usability and utility is essential because it ensures that the developed technologies are not only user-friendly but also practically beneficial in real-world scenarios. 
This echoes a grand challenge for immersive analytics identified by Ens et al.~\cite{ens2021grand} calling for a balanced consideration of usability and utility when evaluating applications aiming to support complex data analysis tasks, rigorously testing novel user interface design and interaction techniques while also reflecting on the practical applications of these designs and techniques. 
Similarly, Billinghurst et al.~\cite{billinghurst:2012} evaluate the educational benefits of immersive analytics applications, demonstrating that while these tools can significantly enhance learning experiences through immersive interactions, their utility depends on the alignment with educational objectives and curriculum needs.

Our work adds to this research with a reflection on usability and the elicitation of use cases following a formative evaluation of our immersive analytics application with participants recruited opportunistically at a large enterprise BI conference.
Our participant pool and their domain expertise are notable, as Friedl-Knirsch et al~\cite{Friedl-Knirsch2024} found that the most prevalent source of participants in evaluations of immersive analytics experiences are students and staff at post-secondary institutions. 
While BI professionals have been the focus demographic of prior formative studies~(e.g.,~\cite{kandel2012enterprise,tory2021finding}), these studies either predate the emergence of immersive analytics or focus on issues relating to desktop-based dashboard applications.


\section{A Formative Evaluation of \app}
\begin{figure}[ht]
 \centering
 \includegraphics[width=\linewidth]{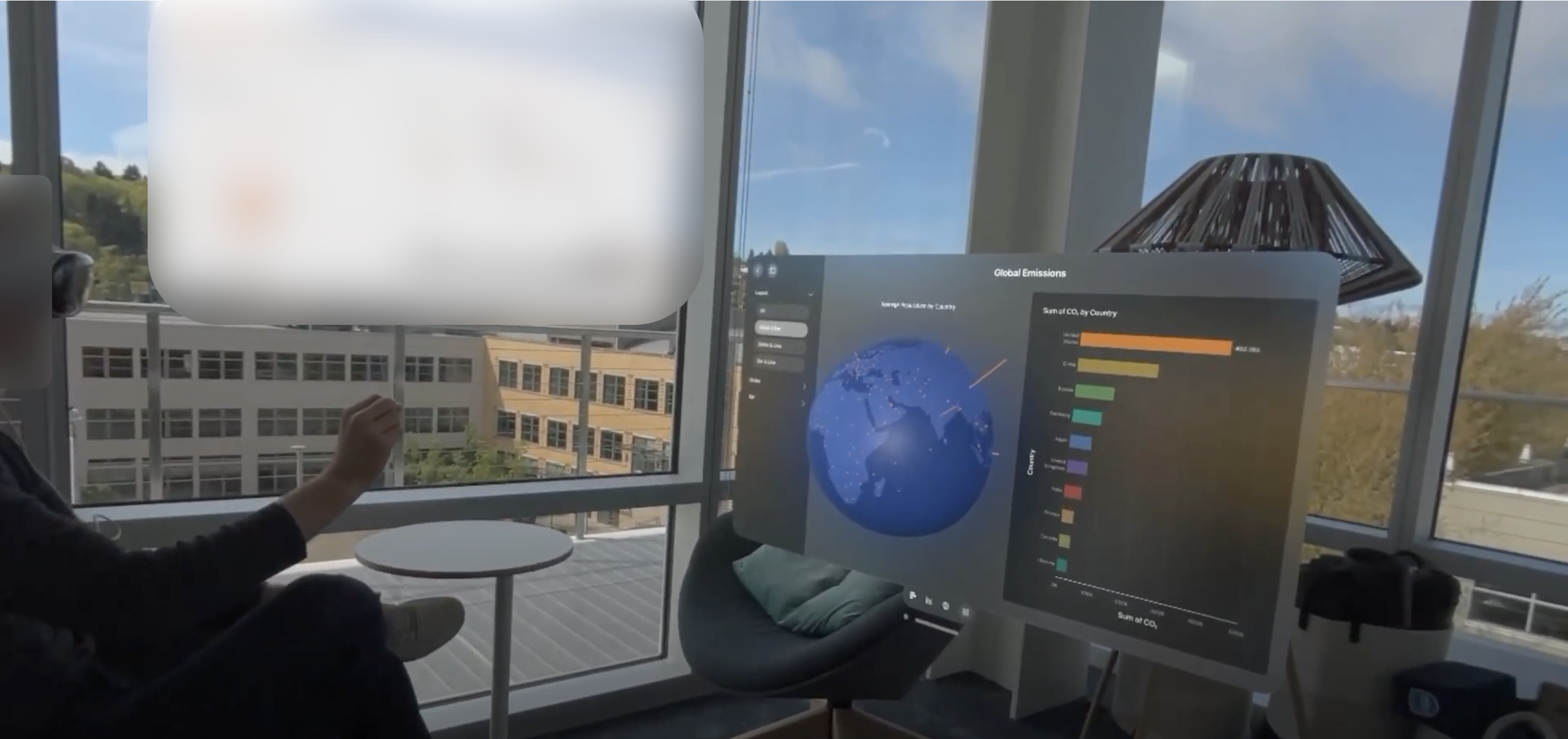}
 \caption{A person wearing an Apple Vision Pro interacts with \app{}. In this instance, a 3D globe and a bar chart display global $CO_2$ emissions data, while a related dashboard appears in a separate web browser window.}
 \label{fig:interface}
\end{figure}

\noindent\textbf{\app{}.}
The design of \app{} as a video-see-through (VST) augmented reality (AR) application followed the general interface design guidelines recommended for Apple's Vision Pro~\cite{visionosguidelines}.
However, as these guidelines are not specific to interacting with data, we also incorporated best practices for the design of interactive visual analytics applications~\cite{few,mackinlayshowme}, such as displaying different visual representations for combinations of data field types and allowing for interactive exploration by applying filters, changing aggregation levels, and switching between alternate representations. 
\app{} leveraged the Vision Pro's spatial computing capabilities for visualizing data in space in proximity to contextually relevant interface elements.  
\app{} featured interactive 2D bar and line charts, as well as 3D globe models for displaying geospatial data (\autoref{fig:interface}). 
\app{} incorporated eye tracking for dwell-based selection as well as hand gestures such as pinching, swiping, and tapping, offering animated visual feedback in response to gesture recognition. 
The application also supported voice-based commands to perform analytical operations such as filtering to a specific country in a dataset. 
Lastly, \app{} supported detail-on-demand interaction~\cite{SHNEIDERMAN2003364} wherein contextual information in the form of tooltips appears when selecting data-bound visual elements.
A demonstration of \app{} is featured in our \href{https://youtu.be/kHQSPnOSpWI&t=1018}{supplemental video}.

\begin{figure}[ht]
  \centering
  \includegraphics[width=\linewidth]{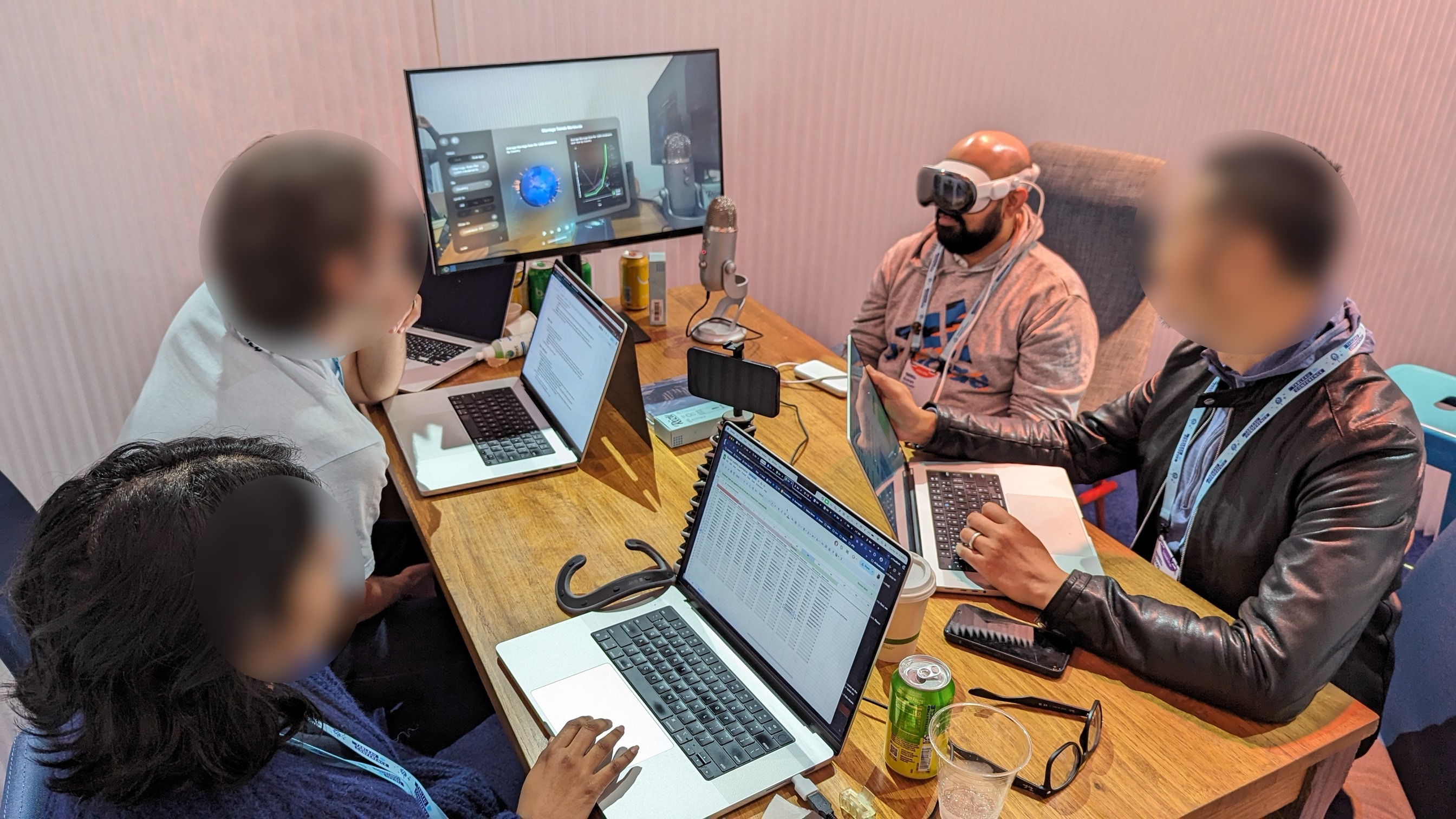}
  \caption{In our evaluation conducted at an enterprise business intelligence conference, participants wearing an Apple Vision Pro head-mounted display (HMD) interacted with \app{} while three study facilitators record their feedback. An external monitor shows a live video feed from the headset. Permission was obtained from the participant shown in the picture.}
  \label{fig:teaser}
\end{figure}

\noindent\textbf{Scope of evaluation.}
Evaluation in HCI is often associated with usability testing via controlled experiments, which typically involve measuring task completion times, error rates, and subjective feedback. 
However, the evaluation of data analytics applications often encompasses a broader spectrum of methodologies tailored to assess the effectiveness of visual data representation for analytical reasoning, work practices, communication, and collaborative data analysis~\cite{lam2011empirical}. 
To this end, we conducted a formative evaluation of \app{} to assess its usability as well as its potential for supporting use cases related to enterprise BI; in other words, we used \app{} as a design probe.  

The evaluation consisted of a series of moderated interviews with 22 participants. 
Study sessions took place over three days at the 2024 Tableau Conference in San Diego, California, which attracted approximately 8,000 Tableau customers and technology partners. We recruited participants at a user research pavilion located in the conference exhibition hall. 
Participant roles included data analysts, managers, and market research consultants. 
The sessions were limited to 45 minutes so as to conform to the conference program schedule, a constraint that was beyond our control. 
While all of our participants had expertise and interest in the domain of BI, expertise in the use of interactive data analytics applications varied: most (12 participants) claimed to be experts, while others reported intermediate expertise (8 participants) or novice status (2 participants). 

\noindent\textbf{Evaluation procedure.}
During each session, one of the authors made qualitative observations while another followed a structured checklist of usability concerns for each feature in \app{}, while another moderated (see \autoref{fig:teaser}). 
The authors identified key findings and themes during daily debriefs and captured them in a collaborative online workspace for later analysis. 

\noindent\textit{Introduction} [$\sim$5min]:
We provided participants with an overview of the study and collected their background information, including their experience with data analytics and familiarity with HMDs.

\noindent\textit{Calibration} [$\sim$5min]:
We guided participants through device fitting and calibrating the device to their eyes and hands, introducing them to the gaze + pinch interaction paradigm of visionOS.

\noindent\textit{Tasks} [$\sim$25min]:
We asked participants to choose one of three demonstration datasets sourced from Our World in Data: livestock production~\cite{livestock}, $CO_2$ emissions~\cite{emissions}, and marriage statistics~\cite{marriage}, with each dataset providing metrics at a national scale.
Together, these datasets are representative of those encountered in many BI contexts, combining quantitative, categorical, temporal, and geospatial attributes.
We then asked participants to perform several tasks involving 2D bar chart and line charts, a 3D globe, and a dashboard containing both 2D charts and a 3D globe. 
Overall, the tasks required specifying filter settings, changing data fields, and interpreting trends. 
Examples of tasks included identifying the country with the highest $CO_2$ emissions in Asia or determining when poultry production first exceeded beef production in South America.

\noindent\textit{Discussion} [$\sim$10min]:
After taking off the device, we asked participants for candid feedback on \app{}'s usability as well as suggestions for new features or improvements.
Then, we treated \app{} like a design probe, in that we asked participants to extrapolate from their initial experience with and consider the use cases, types of data, and visual representations prevalent in their own work that would be amenable to similar immersive analytics experiences.

\section{Reflections and Recommendations}
Our evaluation of \app{} illuminated a tension between usability and the elicitation of use cases, a common challenge when introducing new experiences, such as those involving head-mounted displays. 
This dual focus is particularly pertinent with respect to the Vision Pro, where the novel interaction paradigm of gaze + pinch~\cite{pfeuffer2024design} presented challenges as well as opportunities for rethinking how we evaluate immersive analytics experiences.

\noindent\textbf{Challenge: Assessing both usability and utility.} 
A 45-minute session is admittedly too brief of an interval to investigate both usability and utility for any visual analytics application, and this is especially true of immersive analytics applications.  
However, given the opportunistic nature of conducting an evaluation with BI professionals at an event like the Tableau Conference, as well as the need to satisfy competing stakeholder demands, we nevertheless attempted to report on both aspects.
In reflection, we argue that the elicitation of immersive analytics use cases is at odds with the assessment of usability and feature discoverability. 
For many of the participants, our study not only marked their first time performing gaze + pinch interaction, but it was also their first time interacting with data represented as a 3D object (\i.e.,~the 3D globe displaying geospatial data). We expect that learning how to interact with these data elements detracted from participants' ability to articulate how data in their day-to-day work could be similarly represented.  
It is also worth remarking that a level of secrecy imposed on \app{}'s development precluded a typical iterative human-centered design process with enterprise BI professionals leading up to this evaluation. 
This secrecy is not atypical of enterprise software development, particularly when associated with emerging technologies, and was undoubtedly a factor in our decision to treat \app{} as a starting point for eliciting use cases. 

\textit{Recommendation.} Future evaluations of immersive analytics applications with enterprise professionals should incorporate a combination of usability-focused and heuristic evaluations. The former might involve detailed task-based assessments where participants perform specific actions, followed by debriefing interviews to gather qualitative insights. The latter could serve to identify design inconsistencies and usability flaws. Meanwhile, assessments of utility and the elicitation of usage scenarios could assume a different format altogether: for example, a researcher might perform a rehearsed demonstration of an immersive analytics application to participants in a small focus group setting, in which the group gathers expertise both in visual analytics applications and in spatial computing / HMD-based applications.

\noindent\textbf{Challenge: The novel allure of spatial computing.}
Although all of our participants used BI software in their day-to-day work, an immersive analytics application was not what consistently drew them to our study. 
Rather, some of our participants admitted a driving curiosity to try the Vision Pro (which, at the time of our study, had only been publicly available in the United States for two months). 
Our interviews with this subset of participants revealed no tangible use case in relation to their day-to-day analytics work or the work of those in their organizations.
This is not altogether surprising, as those who use BI applications on a daily basis (a demographic well-represented at the Tableau Conference) are more likely to spend much of their time working with and wrangling tabular (i.e.,~non-spatial) data~\cite{kandel2012enterprise} and maintaining dashboards~\cite{tory2021finding} for others within their organization to use. 
What was surprising, however, was that our participants generally found it difficult to imagine others in their organizations using an immersive analytics application.
For example, an executive might interact with a BI dashboard via a desktop or mobile app, but the prospect of an executive regularly donning a Vision Pro to interact with an immersive BI experience seems unlikely.
Skepticism with respect to immersive BI experiences for executives could be attributed in part due to the cumbersome form factor of most HMDs, and in part due to the (predominantly) non-spatial nature of BI data, which does not typically warrant 3D representation.   
On the other hand, participants were more receptive to the idea of synchronous collaboration support that would allow for multiple people to discuss BI artifacts in virtual spaces, where each conversant would to be able to interact with the same representations of data at the same time from different locations. 

\textit{Recommendation.} A professional interest in data analytics or BI is insufficient when recruiting participants to evaluate the potential utility of an immersive analytics application. A sense of novelty with respect to the Vision Pro and similarly capable HMDs will diminish over time, so the issue of participants being merely curious about the technology or medium may resolve itself. We recommend that prospective study participants volunteer an immersive analytics use case in a short-form text entry as part of a screening survey, as this could help researchers identify participants who are more aligned with the study's objectives.

\noindent\textbf{Challenge: Accounting for participants' expertise.} 
Our observations revealed a dichotomy in participant experiences between those who self-identified as experts in data analytics software and those who reported to be novices. Experts adapted quickly and discovered more features with less prompting. They found our application's workflow to be familiar and aligned well with existing mental models. In contrast, novices required additional guidance and struggled with some of the advanced interaction paradigms. 
Further complicating this difference in performance was the varied level of familiarity with HMDs across both groups, and a subset of the participants had seen a recording of a highly-rehearsed usage demonstration of our application in an \href{https://youtu.be/kHQSPnOSpWI}{earlier session at the Tableau Conference}.


\textit{Recommendation.} Recruiting participants for evaluations of immersive analytics applications perhaps requires a greater level of attention and care relative to studies evaluating conventional desktop-based visual analytics applications. 
Procedures and data collection goals may need to be adapted when engaging with participants with varied levels of expertise with respect to both visual analytics and prior use of immersive applications / HMDs. 
This could require producing demonstrations of varying lengths and levels of detail for different expertise levels.
Those with higher self-reported expertise in both may be better suited to comment on the application's utility, while those with mixed or novice levels of expertise may be better suited to performing tasks that allow researchers to assess the application's learnability and usability.

\noindent\textbf{Limitation: Evaluators were not similarly immersed.} 
We could not directly share participants' experiences while wearing the HMD, but we were able to cast a flattened first-person perspective to a laptop and external display (see \autoref{fig:teaser}) via Apple's AirPlay protocol. 
However, the reliability and fidelity of this video are not analogous to watching a person interact with a conventional desktop-based visual analytics application, particularly given the indirect manipulation of virtual objects with gaze + pinch interaction in visionOS.  
Following the example of James et al. ~\cite{james2023evaluating}'s study in which both the study moderator and participant wore a HoloLens optical see-through HMD, it would be useful for a study moderator to join the participant in a common immersive context wearing their own HMD. At the time of our study (May 2024), visionOS did not yet support co-located collaboration (announced in mid-2025). Alternatively, a moderator could observe a participant via a FaceTime call from from another location, appearing as a spatial persona if using a Vision Pro.

In designing our study, we discovered that visionOS disallows first-person recording while simultaneously casting to an external display. 
Due to this limitation, we recorded a video of the participant from a stationary camera in the room. 
In hindsight, these recordings were of little practical use, as the video did not capture elements of the virtual scene, and the audio quality was poor due to ambient noise. However, it is not clear that a video from the participant's first-person perspective would be beneficial to researchers. First-person perspective video must be stable so as not to induce motion sickness when viewed later, and we cannot expect participants, especially those who are new to wearing HMDs, to be stable camera persons while performing study tasks. Assuming that a study moderator is also wearing an HMD and can observe the participants as they interact, we could record from the moderator headset, as the video would capture participants' interaction more effectively.


\section{Conclusion \& Future Work}
The evaluation of \app{} revealed both the potential and the challenges of evaluating immersive analytics applications with enterprise BI professionals. 
Our study highlights the tension between evaluating usability and eliciting use cases, emphasizing the need for designing appropriate evaluation techniques. 

Additionally, addressing challenges related to capturing the participant experience and overcoming participants' novelty bias is crucial for obtaining useful feedback on the utility of such applications and for eliciting viable analytics workflows. 
Our experience also leads us to anticipate the need for research into the adoption, long-term engagement, and retention strategies for immersive analytics applications in enterprise BI contexts. Diary-based studies and experience sampling protocols may need to be rethought in an immersive context, and researchers may need to consider how to meet with participants remotely in shared virtual environments.
Aggregating and analyzing multiple participants' experiences using approaches such as those used to evaluate Hubenschmid et al.'s ReLive~\cite{hubenschmid2022relive} could also reveal usability and learnability issues.
Finally, given our participants' receptivity to potential collaborative use cases, future research should also explore long-term engagement and retention strategies for immersive analytics tools in the context of synchronous team-based data analysis and decision-making.

To support the further evaluation of immersive analytics applications like \app{} in enterprise BI contexts, our research community must develop new evaluation techniques appropriate for this emerging medium.
Conventional usability testing methods may not fully capture the complexities and nuances of immersive analytics applications. For example, evaluation techniques should account for the unique affordances and challenges of HMDs, such as spatial interaction patterns, real-time data manipulation, and continuous experiences between different devices and platforms. Developing comprehensive frameworks that integrate both qualitative and quantitative measures will be essential to gaining a holistic understanding of experiences in this emerging medium. 

\section*{Supplemental Material}
A video demonstration of \app{} is available at \href{https://youtu.be/kHQSPnOSpWI&t=1018}{youtu.be/kHQSPnOSpWI} (beginning at 16m:58s). We also deployed a \href{https://web.archive.org/web/20241013223759/https://testflight.apple.com/join/QuGwIUHk}{beta version of \app{} to Apple's TestFlight service: testflight.apple.com/join/QuGwIUHk}.

\acknowledgments{%
All of the authors conducted this research while at Tableau.
}

\bibliographystyle{abbrv-doi-hyperref}
\bibliography{main}




\end{document}